\documentclass[prb,floatfix,twocolumn,showpacs,amsmath,amssymb]{revtex4}

\usepackage{graphicx}
\usepackage{dcolumn}

\begin{document}

\title{Finite-temperature properties of frustrated classical spins coupled to the lattice}

\author{C\'edric Weber,$^{1,2}$ Federico Becca,$^{3}$ and Fr\'ed\'eric Mila$^{1}$}
\affiliation{
$^{1}$ Institute of Theoretical Physics, Ecole Polytechnique F\'ed\'erale 
de Lausanne, BSP, CH-1015 Lausanne, Switzerland \\
$^{2}$ Institut Romand de Recherche num\'erique sur les Mat\'eriaux, 
Ecole Polytechnique F\'ed\'erale de Lausanne, CH-1015 Lausanne, Switzerland \\
${^3}$ INFM-Democritos, National Simulation Centre, and 
International School for Advanced Studies (SISSA), I-34014 Trieste, Italy.
}

\date{\today}

\begin{abstract}
We present extensive Monte Carlo simulations for a classical antiferromagnetic 
Heisenberg model with both nearest ($J_1$) and next-nearest ($J_2$) exchange 
couplings on the square lattice coupled to the lattice degrees of freedom.
The Ising-like phase transition, that appears for $J_2/J_1>1/2$ in the 
pure spin model, is strengthened by the spin-lattice coupling,
and is accompanied by a lattice deformation from a tetragonal symmetry to an 
orthorhombic one.
Evidences that the universality class of the transition does not change with
the inclusion of the spin-lattice coupling are reported.
Implications for ${\rm Li_2VOSiO_4}$, the prototype for a layered $J_1{-}J_2$
model in the collinear regime, are also discussed.
\end{abstract}


\maketitle

\section{introduction}\label{intro}

The role of frustrating interactions in low-dimensional systems is a very 
important aspect in the modern theory of magnetism in solids~\cite{book} 
and molecular clusters.~\cite{molecular} 
In particular, the presence of finite-temperature phase 
transitions in two-dimensional systems with 
continuous spin-rotational symmetry, that are ruled out by a naive
interpretation of the Mermin and Wagner theorem, has been clearly documented 
during the last years and continues to attract much attention, because of the 
variety of phenomena that could be generated at low 
temperature.~\cite{henley,ccl,loison,singh,cedric,caprio,sachdev}
The main point is that the presence of frustrating interactions can induce 
non-trivial degrees of freedom, that may undergo a phase transition when the 
temperature is lowered. 
Among the strongly frustrated spin systems, probably one of the most
important examples is the antiferromagnetic Heisenberg model on the square 
lattice with both nearest ($J_1$) and next-nearest neighbor ($J_2$) couplings,
for which it has recently become certain that a very interesting 
scenario shows up for large enough frustrating ratio $J_2/J_1$.
Indeed, for $J_2/J_1>1/2$ and classical spins, the two sublattices are 
completely decoupled at zero temperature,
and each of them has an independent antiferromagnetic 
order. Therefore, the ground-state energy does not depend on the relative 
orientation between the magnetizations of the two sublattices and the 
ground state has an O(3)$\times$O(3) symmetry.
At low temperature, thermal fluctuations lift this huge degeneracy
by an {\it order by disorder} mechanism~\cite{villain} and two families of
collinear states are entropically selected, with pitch vectors 
${\bf Q}=(0,\pi)$ and $(\pi,0)$, respectively.
Chandra, Coleman, and Larkin~\cite{ccl} argued that this fact 
reduces the symmetry to O(3)$\times Z_2$, and that the $Z_2$ symmetry is broken
at low temperature, giving rise to an Ising-like phase transition at finite 
temperature.
Recently,~\cite{cedric} it has been possible to verify 
this scenario by using extensive Monte Carlo simulations,
and a rather accurate estimate of the critical temperature as a function
of the ratio $J_2/J_1$ has been obtained.

The fact that the two states with ${\bf Q}=(0,\pi)$ and $(\pi,0)$
break the rotational symmetry, having antiferromagnetic spin correlations in
one spatial direction and ferromagnetic correlations in the other, 
suggests that, once the spin-lattice coupling is considered,
the lattice could also  experience a phase transition,
ferromagnetic and antiferromagnetic bonds acquiring different lengths.
Indeed, on general grounds, the super-exchange couplings come from a 
virtual hopping of the electrons, which depends upon the actual distance
of the ions.
A first evidence that a similar distortion of the lattice appears
when the spins are coupled to classical lattice distortions has been found 
in Ref.~\onlinecite{becca}, where the quantum $J_1{-}J_2$ model at zero 
temperature has been analyzed by Lanczos diagonalization on small clusters.
In particular, by considering a $4 \times 4$ lattice and spin-1/2 coupled 
to adiabatic phonons, it has been shown that there is a large region where 
the lattice distorts and ferromagnetic and antiferromagnetic bonds acquire 
different lengths.

Interestingly, there is also clear evidence that such a scenario is realized in 
a real compound, ${\rm Li_2VOSiO_4}$, a vanadate which can be considered as
a prototype of the $J_1{-}J_2$ model in the collinear 
region:~\cite{millet,carretta}
Indeed, although the value of $J_2/J_1$ is not exactly
known,~\cite{carretta,rosner,misguich} all estimates indicate that
$J_2 \gtrsim J_1$. Moreover, NMR and muon spin rotation magnetization
provide a clear evidence for the presence of a phase transition to a collinear 
order at $T_c \sim 2.8 K$ and a structural distortion at a nearby temperature.
A simple and appealing explanation relies on the existence of the Ising-like
transition and the concomitant lattice distortions.

Since an unbiased finite-temperature analysis of the quantum model for 
large clusters is, at present, impossible, in this work we would like to 
address the simpler problem of classical spins coupled to lattice distortions.
Although this could be viewed as a crude approximation to the original 
quantum model, especially for low temperature and near the disordered 
region (expected near $J_2/J_1 \sim 1/2$), the classical
case of spins coupled to lattice deformation represents a highly non-trivial 
problem and, certainly, gives the zeroth-order approximation for the true 
quantum case.

The paper is organized as follows: In Sec.~\ref{sec:model}, we present the
model and the method we used and in Sec.~\ref{sec:results} we show the 
results and draw the conclusions.

\section{The model and the method}\label{sec:model}

In this section we introduce the spin-lattice Hamiltonian and briefly 
describe the method we used to treat the lattice degrees of freedom.
The Hamiltonian reads:
\begin{eqnarray}\label{ham}
&&\hat{\cal{H}} = \sum_{\langle i,j \rangle} J_1(d_{ij})
\hat{{\bf {S}}}_{i} \cdot \hat{{\bf {S}}}_{j}
+ \sum_{\langle \langle i,j \rangle \rangle} J_2(d_{ij})
\hat{{\bf {S}}}_{i} \cdot \hat{{\bf {S}}}_{j} + \nonumber \\
&& \frac{K_1}{2} \sum_{\langle i,j \rangle} 
\left ( \frac{d_{ij}-d_{ij}^0}{d_{ij}^0} \right )^2
+ \frac{K_2}{2} \sum_{\langle \langle i,j \rangle \rangle} 
\left ( \frac{d_{ij}-d_{ij}^0}{d_{ij}^0} \right )^2,
\end{eqnarray}
where $\hat{{\bf {S}}}_{i}$ are O(3) spins on a periodic square lattice 
with $N=L \times L$ sites. $\langle i,j \rangle$ and 
$\langle \langle i,j \rangle \rangle$ indicate the sum over nearest and 
next-nearest neighbors, respectively. The super-exchange couplings 
$J_1(d_{ij})$ and $J_2(d_{ij})$ depend upon the distance $d_{ij}$ 
of the sites $i$ and $j$, and $K_1$ and $K_2$ are the elastic coupling
constants. Finally, $d_{ij}^0$ is the {\it bare} lattice distance.
In transition metal compounds, general arguments lead to exchange integrals
that vary like the inverse of the distance to a certain power $\theta$,
with $\theta$ in the range $5{-}15$.~\cite{harrison} Therefore, for small 
displacements around the equilibrium positions, we can write:
\begin{equation}\label{jj}
J(d_{ij})=J \left ( \frac{d_{ij}^0}{d_{ij}} \right )^\theta \simeq
J \left [ 1 - \theta \left ( \frac{d_{ij} - d_{ij}^0}{d_{ij}^{0}} \right )
\right ]~.
\end{equation}

In the Hamiltonian of Eq.~(\ref{ham}), we treat both the spins 
$\hat{{\bf {S}}}_{i}$ and the lattice coordinates $d_{ij}$ as dynamical
variables, allowing them to change their configurations. 
With respect to the pure spin problem, this fact doubles the number of 
dynamical variables present in the problem and makes the Monte Carlo algorithm
much heavier. In particular, we find better to use Monte Carlo algorithms 
with local and/or global updates, instead of more involved Monte Carlo methods
based on the reconstruction of the density of states,~\cite{brazilian,wl} 
that we used in the case where the lattice is kept fixed.

In practice, we work on a torus with two {\it independent} lengths $L_x$ and 
$L_y$, which play the role of additional degrees of freedom sampled by Monte
Carlo. Moreover, each site of the cluster can independently change its position 
in the lattice. These facts allow us to
consider structural distortions as well as the usual spin properties.
We found that, besides small displacements around their equilibrium positions,
the sites always form either a square cluster, with tetragonal symmetry and
equal lattice spacing, $l_x$ and $l_y$, in the two spatial directions 
(i.e., $l_x=l_y$) or a rectangular cluster, with orthorhombic symmetry and 
different lattice spacings in the two directions (i.e., $l_x \ne l_y$).
In these cases the total lengths of the cluster in the two directions are 
$L_x = L \times l_x$ and $L_y = L \times l_y$. 

For the pure spin model, e.g., with fixed distances and 
$J(d_{ij})=J$, in order to characterize the Ising-like phase transition, 
it is useful to construct, from the original spin variables 
$\hat{{\bf {S}}}_{i}$, an effective Ising variable 
(on the dual lattice):~\cite{ccl}
\begin{equation}\label{isingvar}
\sigma_x =
\frac{(\hat{{\bf {S}}}_{i} - \hat{{\bf {S}}}_{k}) \cdot
(\hat{{\bf {S}}}_{j} - \hat{{\bf {S}}}_{l})}
{\vert (\hat{{\bf {S}}}_{i} - \hat{{\bf {S}}}_{k}) \cdot
(\hat{{\bf {S}}}_{j} - \hat{{\bf {S}}}_{l}) \vert}~,
\end{equation}
where $(i,j,k,l)$ are the corners with diagonal $(i,k)$ and $(j,l)$ of the
plaquette centered at the site $x$ of the dual lattice.
In this way, the two collinear states with ${\bf Q}=(\pi,0)$ and 
${\bf Q}=(0,\pi)$ can be distinguished by the value of the
Ising variable, $\sigma_x= \pm 1$.
As emphasized in Ref.~\onlinecite{cedric}, the normalization term does not 
affect the critical properties of the model and it is only introduced to have a
normalized variable. 

\begin{figure}
\vspace{-0.3cm}
\includegraphics[width=0.45\textwidth]{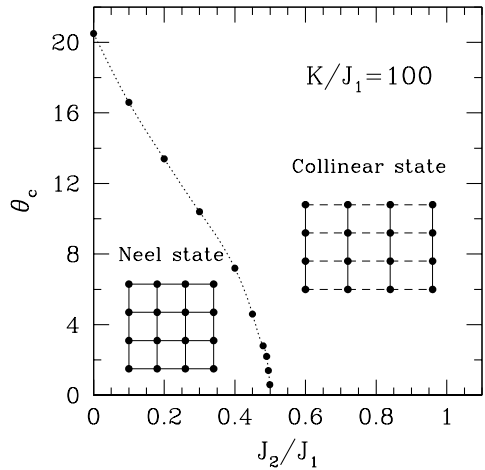}
\vspace{-0.3cm}
\caption{\label{fig:phasediag}
Zero-temperature phase diagram of the Hamiltonian of Eq.~(\ref{ham}): the
N\'eel state, with a tetragonal lattice, is stabilized for $J_2/J_1 < 0.5$ 
and below a critical value of the spin-lattice coupling $\theta_c$, 
otherwise the ground state has collinear spins, with antiferromagnetic
correlations along one spatial direction and ferromagnetic along the other,
and the lattice has an orthorhombic structure. Lines are guides to the eye.
}
\end{figure}

Moreover, we can easily construct the Ising-like order parameter as 
$M_\sigma = (1/N)\sum_x \sigma_x$.
Associated to the Ising magnetization $M_\sigma$, we can define the
susceptibility related to the Ising variable
$\chi=(N/T) (\langle M_\sigma^2 \rangle - \langle |M_\sigma| \rangle^2)$.
Finally, in order to study the finite-temperature phase diagram, it is also
useful to consider the specific heat per site
$C_v=(1/L T^2) (\langle E^2 \rangle - \langle E \rangle^2)$, $E$ being
the total energy of the system. 

Before considering the Monte Carlo results at finite temperature, it is
useful to discuss the zero-temperature phase diagram of the spin-lattice 
Hamiltonian of Eq.~(\ref{ham}), see Fig.~\ref{fig:phasediag}.
In this case, the first-order phase transition present for $\theta=0$, 
i.e., for the pure spin model, at $J_2/J_1=0.5$ bends towards smaller values 
of the frustrating ratio $J_2/J_1$ when the coupling to the lattice is 
switched on.
In particular, the huge 
O(3)$\times$O(3) degeneracy is already broken at zero temperature, 
and the collinear states with pitch vector ${\bf Q}=(\pi,0)$ and 
${\bf Q}=(0,\pi)$ have a lower energy.
It is worth noting that this is a general property of the spin-lattice
coupling, which is also present in the quantum case.~\cite{becca}
Therefore, on the basis of the analysis presented in Ref.~\onlinecite{cedric},
in the presence of the spin-lattice coupling, we expect that
a finite-temperature phase transition shows up also for a {\it bare} ratio
$J_2/J_1<1/2$, whenever the coupling $\theta$ is sufficiently large,
i.e., $\theta > \theta_c$. In the following, we will present strong 
evidences that there is a finite-temperature phase transition
that is related to a change in the lattice symmetry, from a high-temperature
disordered magnetic phase with a tetragonal lattice to a low-temperature
Ising-ordered
phase, with $M_\sigma \ne 0$, and with an orthorhombic lattice.
A little bit more subtle is the determination of the universality class of
this transition. By using powerful Monte Carlo methods, we showed that, 
in the absence of spin-lattice coupling, the transition belongs to the 
Ising universality class,~\cite{cedric} as expected from more general 
arguments.~\cite{ccl}
The concomitant presence of spin and lattice degrees of freedom makes the
Monte Carlo calculation much heavier than in the simpler case where only
spins are considered, and we are limited to smaller clusters. Nevertheless,
we can reach rather large lattices, that enable us to have convincing
results for the thermodynamic limit.

\section{Results and discussion}\label{sec:results}

In this section, we present the Monte Carlo results for the Hamiltonian of 
Eq.~(\ref{ham}), and we give evidences that the Ising transition present in
the pure spin model survives when the spin-lattice coupling is taken into 
account, and, moreover, that it is accompanied by a structural deformation. 

As discussed in the previous section, in the presence of both spin and
lattice degrees of freedom, the Monte Carlo simulations become rather expensive
and we are limited concerning the size of the clusters. In order to have a 
sufficiently high transition temperature $T_c$, that allows a reliable 
estimate of the physical properties, most of the calculations have been 
performed for $J_2/J_1 = 0.8$ and three different values of the spin-lattice 
coupling, $\theta=5$, $10$, and $15$ for $K_1=K_2=K$. In the following we 
consider the case of $K/J_1=100$.
Then, we also report calculations for other values of the ratio $J_2/J_1$,
where a similar behavior is observed.
Therefore, on the basis of our numerical results, we argue that different 
choices for the frustrating ratio $J_2/J_1$ or for the elastic coupling $K_1$ 
and $K_2$ do not change qualitatively the physical behavior of the system.
However, it should be emphasized that a precise determination of the 
critical exponents could be difficult and, in general, rather large clusters 
are needed. In the following, we present evidences that the transition 
belongs to the Ising universality class, which has been clearly documented 
in the absence of spin-lattice coupling.

\begin{figure}
\vspace{-0.3cm}
\includegraphics[width=0.45\textwidth]{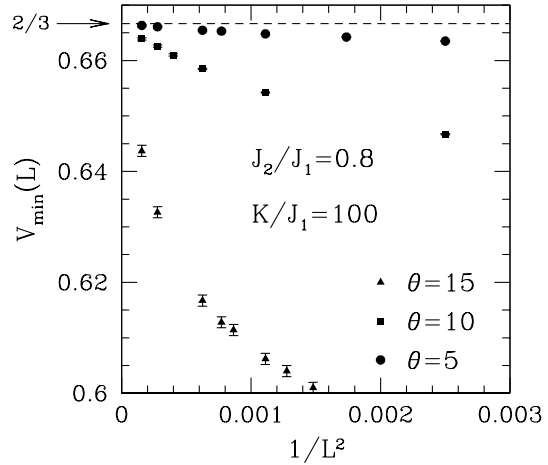}
\vspace{-0.3cm}
\caption{\label{fig:U4}
$V_{\min }(L)$ as a function of $1/L^2$ for $\theta=5$, $10$, and $15$. 
The horizontal line, located at $2/3$, indicates the value for a continuous 
phase transition.
}
\end{figure}

\begin{figure}
\vspace{-0.3cm}
\includegraphics[width=0.45\textwidth]{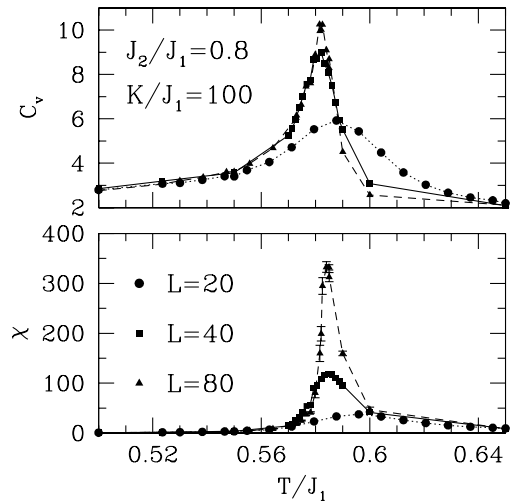}
\vspace{-0.3cm}
\caption{\label{fig:chia5}
Upper panel: Specific heat as a function of the temperature for $\theta=5$
for three different clusters $L=20$, $40$, and $80$. 
Lower panel: The same for the spin susceptibility.
}
\end{figure}

\begin{figure}
\vspace{-0.3cm}
\includegraphics[width=0.45\textwidth]{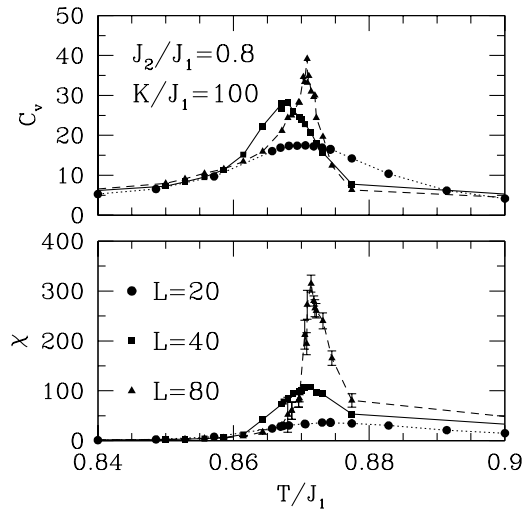}
\vspace{-0.3cm}
\caption{\label{fig:chia10}
The same as in Fig.~\ref{fig:chia5} for $\theta=10$.
}
\end{figure}

First of all, in order to have some useful insight into the order of the 
phase transition, we consider Binder's fourth energy cumulant:
\begin{equation}
V_L(T) = 1 - \frac{\langle {E^4 } \rangle }{3 \langle {E^2 } \rangle }.
\end{equation}
Indeed, by considering the size scaling of $V_{min}(L)$, the value of 
$V_L(T)$ at the ``critical'' temperature of the finite cluster $T_c(L)$,
it is possible to discriminate between a continuous transition, for which
$V^* = \lim_{L \to \infty} V_{min}(L) = 2/3$, and a first-order transition,
for which $V^*<2/3$.~\cite{binder}
The size dependence of  $V_L(T_c)$ is shown in Fig.~\ref{fig:U4} for three 
values of $\theta$. Interestingly, in all the cases considered 
$V_{min}(L) \to 2/3$, giving a clear evidence that the transition is
continuous and not first order.

In Figs.~\ref{fig:chia5} and~\ref{fig:chia10}, we show the specific heat 
and the spin susceptibility $\chi$ for $\theta=5$ and $10$, respectively.
In both quantities, the presence of a huge peak, that grows with the 
size of the cluster, marks the presence of a phase transition.
It is worth noting that, while the maximum value of the spin susceptibility 
is not affected by the spin-lattice coupling, the maximum value of the 
specific heat strongly depends upon $\theta$. 
Moreover, the low-temperature value of the specific
heat is renormalized by the presence of the lattice displacements: the
bare value $C_v=1$, that holds when the lattice is frozen in its equilibrium
configuration, is changed into $C_v=2$ whenever the sites can have small 
displacements around their equilibrium positions.

\begin{figure}
\vspace{-0.3cm}
\includegraphics[width=0.45\textwidth]{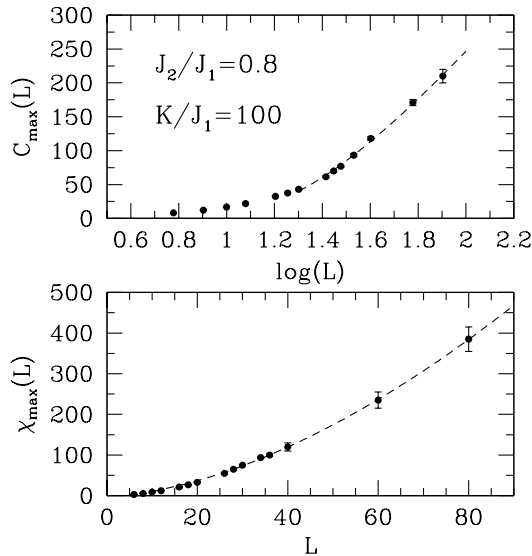}
\vspace{-0.3cm}
\caption{\label{fig:scaling}
Upper panel: Size scaling of the maximum of the specific heat as a
function of the size of the cluster for $\theta=15$.
Lower panel: The same for the maximum of the spin susceptibility.
In both cases, the dashed line is a three-parameter fit (see text).
}
\end{figure}

The actual determination of the universality class of the transition can be
assessed by considering the size scaling of the physical quantities. 
For a second-order phase transition, the maximum value of the susceptibility 
and of the specific heat follow well defined scaling relations, i.e.,
$\chi_{max}(L) \sim L^{\gamma/\nu}$ and 
$C_{max}(L) \sim L^{\alpha}$ (a logarithmic behavior, i.e., 
$\alpha=0$, for the Ising universality class).
On the other hand, for a discontinuous phase transition, we have
$\chi_{max}(L) \sim L^2$ and $C_{max}(L) \sim L^2$.
As clearly shown for $\theta=0$,~\cite{cedric} when the spins are completely
decoupled from the underlying lattice, the phase transition is second order
and falls into the Ising universality class, i.e., $\alpha=0$, $\nu=1$ and
$\gamma=7/4$. In this case, an accurate determination of $\alpha$ is quite 
hard and rather large clusters are needed (i.e., $L \sim 200$). 
In the case of a finite $\theta$, we cannot afford such large clusters, but,
nevertheless, we can obtain rather convincing evidences in favor of the Ising 
universality class by considering the case of a large spin-lattice 
coupling $\theta$.
Indeed, in Fig.~\ref{fig:scaling}, we show the size scaling of the maximum 
of the specific heat and of the spin susceptibility for $\theta=15$. 
As mentioned before, the value of the peak of the spin susceptibility does 
not depend upon $\theta$, and, therefore, the results are also valid for other
values of the spin-lattice couplings.
A three-parameter fit of $\chi_{max}(L)=a + b \times L^{\gamma/\nu}$ gives 
$\gamma/\nu =1.7 \pm 0.1$, which is in reasonable agreement with the value 
$1.75$ of the Ising transition. 
Moreover, the clearest evidence that the transition belongs to the Ising
universality class comes from the specific heat. Indeed, the best fit of the 
maximum of the specific heat is given by
$C_{max}(L)= a_0 + a_1 \log(L) +a_2/L$,~\cite{ferdinand} while a power
law turns out to be completely inadequate.
This fact completely rules out the possibility to have a first-order
phase transition, for which a power law with an exponent $\alpha=2$ is 
expected.
For smaller values of $\theta$ it is much harder to get an accurate
fit of $C_{max}(L)$, and much larger sizes should be considered.
Nevertheless, having in mind the results for $\theta=0$, we expect that
the phase transition also falls into the Ising 
universality class for small spin-lattice coupling.~\cite{note}

\begin{figure}
\vspace{-0.3cm}
\includegraphics[width=0.45\textwidth]{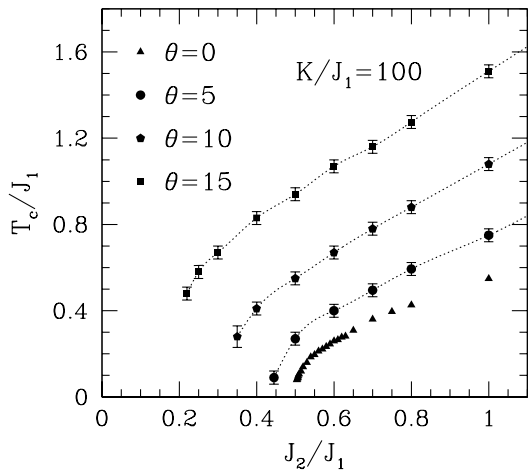}
\vspace{-0.3cm}
\caption{\label{fig:tc}
The critical temperature $T_c$ as a function of the frustrating ratio
$J_2/J_1$ for three different values of the spin-lattice couplings,
$\theta=5$, $10$, and $15$. The case for the pure spin model, corresponding 
to $\theta=0$, is also reported for comparison.
}
\end{figure}

By using a similar analysis for different $J_2/J_1$, it is possible to 
determine the behavior of the critical temperature $T_c$ as a function of 
the frustrating ratio $J_2/J_1$ for different values of $\theta$. 
The 
results are reported in Fig.~\ref{fig:tc}.
In particular, in order to determine $T_c$, we used two alternative methods 
that give consistent results: On one hand, we take the temperatures for 
which the susceptibility has its maximum and we make use of the scaling 
$T_c(L) \sim T_c + a \times L^{-1}$ 
(which assumes $\nu=1$); on the other hand, we used Binder's fourth cumulant
of the Ising parameter. 
Interestingly, as already anticipated, for a sufficiently large value of the 
spin-lattice coupling there is a phase transition to a collinear phase even if
$J_2/J_1<1/2$ as soon as the spin-lattice coupling is switched on,
the critical ratio $J_2/J_1$ for which a transition occurs for a given
$\theta$ being in good agreement with the static analysis 
(Fig. \ref{fig:phasediag}).

\begin{figure}
\vspace{-0.3cm}
\includegraphics[width=0.45\textwidth]{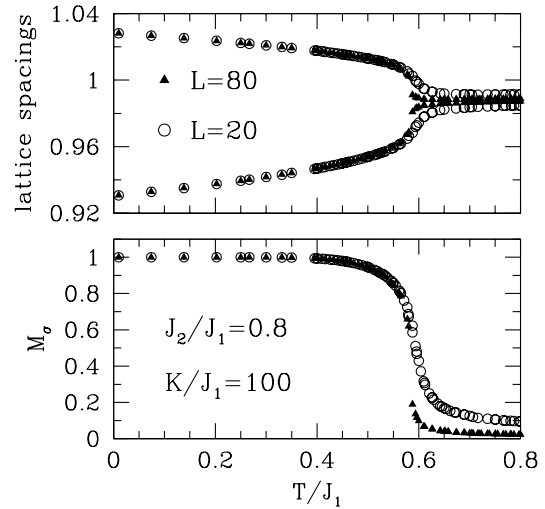}
\vspace{-0.3cm}
\caption{\label{fig:displa5}
Upper panel: lattice parameters $l_x$ and $l_y$ as a function of the 
temperature for $\theta=5$ for $L=20$ and $80$.
Lower panel: The same for the Ising magnetization $M_\sigma$.
}
\end{figure}

\begin{figure}
\vspace{-0.3cm}
\includegraphics[width=0.45\textwidth]{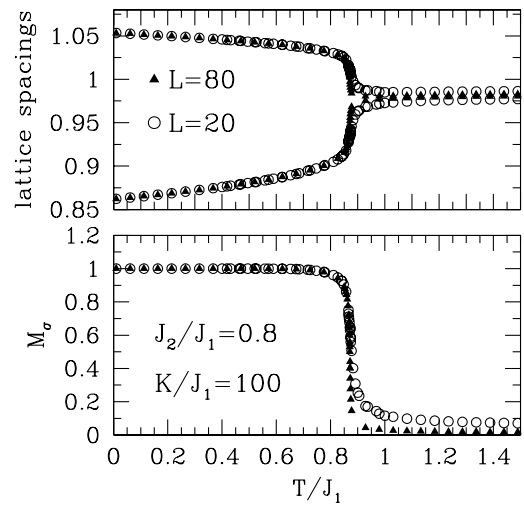}
\vspace{-0.3cm}
\caption{\label{fig:displa10}
The same as in Fig.~\ref{fig:displa5} for $\theta=10$.
}
\end{figure}

Now, we turn to the discussion of the lattice properties.
Because the two families of states with pitch vectors ${\bf Q}=(0,\pi)$ 
and $(\pi,0)$, that are entropically selected, have different spin correlations 
in the two spatial directions, whenever the sites of the lattices are coupled 
to the spins, a structural transition is also likely to show up.
Indeed, while at high temperature the lattice has a 
tetragonal symmetry, with the same lattice spacing in the $x$ and $y$
directions, at the transition temperature, the lattice undergoes a 
structural phase transition towards an orthorhombic symmetry, with
different lattice spacings $l_x$ and $l_y$,
see Figs.~\ref{fig:displa5} and~\ref{fig:displa10}.
This is exactly equivalent to what happens at zero temperature in the
quantum case, where the collinear phase is also accompanied by
a lattice distortion.
Notice that, since we are considering classical spins, 
the only relevant distortion is the orthorhombic one, and, indeed, 
we do not find any evidence towards other displacements of the underlying
lattice, in contrast to the spin 1/2 case.~\cite{becca}

One important outcome is that the structural phase transition has a huge
effect just below the critical temperature. Indeed, the lattice displacements
strongly stabilize one of the two families of states, and almost freeze the
spin fluctuations. This can be clearly seen from the Ising-like magnetization 
$M_\sigma$, whose value is practically fixed to its saturation value from zero 
temperature up to $T_c$, where it goes to zero (or to a very small 
value for finite sizes), see Figs.~\ref{fig:displa5} and~\ref{fig:displa10}.

The implications for quasi-two-dimensional models with inter-plane
magnetic couplings will presumably depend on the value of this coupling.
If it is sufficiently weak, we expect the system to undergo two phase transitions:
First, coming from high temperature, an Ising like transition involving a 
lattice distortion and a breaking of the rotational symmetry of spin 
fluctuations, then a three-dimensional magnetic ordering. However, 
since the correlation length increases exponentially fast at low temperature
in two-dimensional antiferromagnets, the critical temperature for three-dimensional 
ordering is expected to grow very fast with the inter-plane coupling, reaching
quite rapidly values of the order of the intra-plane couplings. In that situation,
the two transitions might very well merge into a single phase transition,
with a critical behavior that could be very different from the standard
one for magnetic ordering of three-dimensional Heisenberg
antiferromagnets though.~\cite{3dheis}
Given the lattice sizes available when phonons are included, this is an issue
we could not address however.

From an experimental point of view, the first prototype of the
spin-$1/2$ $J_1{-}J_2$ model has been recently synthetized.~\cite{millet}
It is a layered vanadium oxide ${\rm Li_2VOSiO_4}$
in which ${\rm VO_5}$ pyramids
are arranged in such a way that second vanadium neighbors are in the same
plane while first neighbors are not, so that $J_2$ need not be
smaller than $J_1$. Although there is no general agreement on the precise 
value of the ratio $J_2/J_1$, both experimental and theoretical estimates
lead to a ratio significantly larger than 1/2.
Therefore, the system is expected to develop collinear order, which
has been confirmed by NMR results at the ${\rm Li}$ site.
Indeed, the NMR spectrum at this site clearly shows that, below 
$T_c \sim 2.8 K$, the central peak splits into three different peaks that 
correspond to different ${\rm Li}$ sites where the local field is either zero 
or nonzero (parallel or antiparallel to the external field).
This behavior has been associated to a magnetic order where the spins lie
along the $x$ direction with the magnetic vector 
${\bf Q}=(0,\pi)$.~\cite{carretta}
Moreover, the $^{29}{\rm Si}$ NMR spectrum is very
anomalous at low temperature,~\cite{carretta} and given the very
symmetric position of ${\rm Si}$ in the lattice, this cannot be attributed
to the development of collinear order but must be related to a
structural phase transition. Given the rather low characteristic temperatures
in that system ($T_c\simeq 2.8 K$ while the transfer of weight of the
${\rm Si}$ line is complete at 2 K), a precise structural determination of the
low temperature phase has not been possible yet. 
Interestingly enough, the temperature at which the system starts to develop
a structural distortion seems to be slightly larger than $T_c$, which 
would agree with the general expectation for small inter-plane coupling.
Indeed, experimentally, it is not possible to exclude that distorted regions 
start to generate above 2.8K and grow, by decreasing the temperature, by a 
nucleation procedure.~\cite{notecarretta}
Whether a more direct determination of the structural transition will 
confirm this point remains to be seen.

\acknowledgments

We are grateful to P. Carretta and L. Capriotti for useful discussions 
throughout this project. One of us (F.B.) acknowledges the warm hospitality 
of EPFL, where this project was initiated. F.B. is supported by COFIN-2004 and 
by INFM. This work was supported by the Swiss National Fund and by MaNEP.


\begin{thebibliography}{99}
\bibitem{note} We cannot exclude that a first-order or a second-order phase
   transition with different exponents appears for some choice of the 
   couplings $\theta$, $K_2/K_1$, or $J_2/J_1$.
\end{thebibliography}
\end{document}